\documentclass[12pt,preprint]{aastex}

\usepackage{graphicx} 
\usepackage{mathptmx}



\def\***#1{{\sc #1}} \def\plan#1{\relax} \def\Plan#1{\relax}
\def\PLAN#1{\relax}

\renewcommand{\arraystretch}{1.2}

\def\lta{\mathrel{\spose{\lower 3pt\hbox{$\mathchar"218$}} \raise
2.0pt\hbox{$\mathchar"13C$}}} \def\gta{\mathrel{\spose{\lower
3pt\hbox{$\mathchar"218$}} \raise 2.0pt\hbox{$\mathchar"13E$}}}
\newcommand{\etal}{{\it et al.}}  

\shortauthors{Jithesh et al.}
\shorttitle{Optically dark X-ray sources}

\usepackage{subfig}

\def\mathnew{\mathsurround=0pt}

\def\simov#1#2{\lower .5pt\vbox{\baselineskip0pt \lineskip-.5pt
\ialign{$\mathnew#1\hfil##\hfil$\crcr#2\crcr\sim\crcr}}}

\def\simless{\mathrel{\mathpalette\simov <}}

\begin{document}

\title{Black Hole Mass Limits for Optically Dark X-ray Bright Sources in Elliptical Galaxies }

\author{V. Jithesh\altaffilmark{1}, K. Jeena\altaffilmark{1}, R. Misra\altaffilmark{2}, S. Ravindranath\altaffilmark{2},
G. C. Dewangan\altaffilmark{2}, C. D. Ravikumar\altaffilmark{1},  and B. R. S. Babu\altaffilmark{1}   }

\altaffiltext{1}{Department of Physics, University of Calicut,
Malappuram-673635, India; jitheshthejus@gmail.com}

\altaffiltext{2}{Inter-University Center for Astronomy and
Astrophysics,  Post Bag 4, Ganeshkhind, Pune-411007, India; rmisra@iucaa.ernet.in
}

\begin{abstract}

Estimation of the black hole mass in bright X-ray sources of nearby
galaxies is crucial to the understanding of these systems and their
formation. However, the present allowed black hole mass range spans
five order of magnitude ($10M_\odot < M < 10^5 M_\odot$) with the
upper limit obtained from dynamical friction arguments. We show that
the absence of a detectable optical counterpart for some of these
sources, can provide a much more stringent upper limit. The argument is
based only on the assumption that the outer regions of their accretion
disks is a standard one. Moreover, such optically dark
X-ray sources cannot be foreground stars or background active galactic nuclei, 
and hence must be accreting systems residing within their host galaxies.
As a demonstration we search for candidates among the point-like X-ray 
sources detected with {\it
Chandra} in thirteen nearby elliptical galaxies. We use a novel
  technique to search for faint optical counterparts in the {\it HST}
  images whereby we subtract the bright galaxy light based on isophotal
  modeling of the surface brightness. We show that for six sources 
with no
detectable optical emission at the 3$-$sigma level, their black hole masses 
$M_{BH} <  5000M_\odot$.
In particular, an ultra-luminous X-ray source (ULX) in NGC~4486 has 
$M_{BH}< 1244 M_\odot$. We discuss the potential of this method to provide
stringent constraints on the black hole masses, and the implications on the
physical nature of these sources.


\end{abstract}


\section{Introduction} 

Compact, off-nuclear X-ray point sources in nearby galaxies, with luminosities
$10^{39}-10^{41}\rm~ergs~s^{-1}$  are
referred to as Ultra-Luminous X-ray sources (ULXs). Detected in the
early 1980's, with the {\it Einstein} X-ray satellite \citep{Fab89},
these objects were further studied with {\it ROSAT} \citep{Col99} and
{\it ASCA} \citep{Mak00}.  The {\it XMM-Newton} and {\it Chandra}
X-ray observatories with their significantly higher angular
resolution, dramatically confirmed the presence of ULXs \citep{Kaa01},
and have enabled their spectral and temporal properties to be studied
in detail \citep[see ][for reviews]{Mil04,Mus04,Mus06,Rob07}.

The observed luminosities of ULXs exceed the Eddington limit
for a $10 M_{\odot}$ black hole.  Since ULX are off-nuclear sources, their masses must be
$< 10^{5} M_{\odot}$ from dynamical friction
arguments \citep{Kaa01}. Thus, ULX may represent a class of
Intermediate Mass Black Holes (IMBHs) whose mass range ($10 M_{\odot}
< M < 10^{5}M_{\odot}$) lies between that of stellar mass black holes
and super-massive black holes observed in galaxy centers
\citep{Mak00}. Alternatively, ULX may be stellar mass black hole
systems exhibiting super-Eddington accretions with their radiation
geometrically beamed \citep{Sha73,Kin08}.

X-ray spectroscopy has provided supporting evidence in favor of IMBHs
of $\sim1000M_{\odot}$ in ULXs \citep{Mil03a,Mil03b,Cro04,Dew04,Rob05,Dev08}. 
Moreover, X-ray timing characteristics, i.e. presence of low frequency
QPOs and/or breaks in the power density spectra, also suggests that ULX may harbor
$\sim 100$-$1000M_{\odot}$ black holes \citep{Str03,Dew06,Muc06,Str09}.
While
indicative these results are not conclusive, since there are also several
arguments against IMBHs in ULXs \citep[see e.g.,][]{Mus04,Rob07}
and further investigations are required to reveal the true nature of
these sources.

Study of the host galaxy properties of ULXs reveals that their number
and total X-ray luminosity is related to recent star formation
activity, suggesting that they originate in young short-lived systems
\citep{Swa04,Swa09}.  While the number of ULXs per
galaxy is roughly the same for both spirals and ellipticals, the ones
in the spirals have significantly higher luminosities \citep{Swa04}.
Optical counterparts have been reported for some ULX
\citep{Liu04,Kun05,Ram06,Ter06}. While some of the counterparts
have been identified as O stars \citep{Liu02,Liu07}, for most ULXs, the optical
counterparts are stellar clusters \citep{Goa02,Pta06}. However, for 
many ULX, the optical counterparts reveal that they are either background AGN \citep{Gut06, Bon09} or foreground stars.
ULXs found in elliptical galaxies may have contamination from 
background sources
 at $\sim 44$\% level \citep{Swa04}. Detailed studies of
X-ray sources in general and their connection with globular clusters have been
undertaken \citep{Kim06, Kim09} who note that the X-ray properties of the
 the sources in the field (i.e. without optical
counterparts) are not different from those in globular clusters.

The allowed black hole
mass range for X-ray sources in nearby galaxies span five orders of magnitude 
($10 M_{\odot} < M < 10^{5}M_{\odot}$) and it is important
to obtain tighter constrains. Here, we show that the absence of a detectable
optical emission allows us to impose an upper 
limit on the black hole mass for these accreting systems based on 
some standard assumptions. Moreoever, We argue that these {\it optically dark} 
X-ray sources cannot be foreground stars or background 
AGN and hence are a true sample of sources located within the host galaxy. 
To demonstrate the technique, we search for candiatates among 
X-ray bright sources detected by {\it Chandra} in archival {\it HST} ACS, 
and WFPC2 images.

\begin{deluxetable} {lccc}
\tablewidth{0pt} 
\tablecaption{Sample Galaxy Properties} 
\tablehead{
\colhead{Galaxy} & \colhead{Distance (Mpc)} &\colhead{$N_{x}$}&\colhead{$N_{d}$}} 
\startdata 

NGC 1399 & $18.3$ & $26$ & $4$ \\
NGC 4649 & $16.6$ & $12$ & $5$  \\
NGC 4697& $11.8$ & $11$ & $3$  \\ 
NGC 1291 & $8.9$ & $5$ & $1$ \\ 
NGC 4365 & $20.9$ & $4$ & $3$  \\
NGC 1316 & $17.0$ & $7$ & $3$  \\
NGC 4125 & $24.2$ & $3$ & $3$ \\  
NGC 3379 & $11.1$ & $3$ & $1$  \\
NGC 4374 & $17.4$ & $2$ & $1$   \\ 
NGC 4486 & $15.8$ & $5$ & $2$   \\ 
NGC 4472 & $15.9$ & $1$ & $0$   \\ 
NGC 1407 & $17.6$ & $2$ & $2$  \\ 
NGC 4552 & $15.9$ & $3$ & $0$  \\ 
\enddata
\tablecomments{(1) Host galaxy name; (2) Distance to the galaxy ; (3) Number 
of X-ray sources within HST field of view; (4) Number of X-ray sources without optical counterparts}
\label{Sample}
\end{deluxetable}

\begin{figure*}
\begin{center}

  \subfloat[NGC 4486]{\label{fig:NGC 4486}\includegraphics[width=0.25\textwidth]{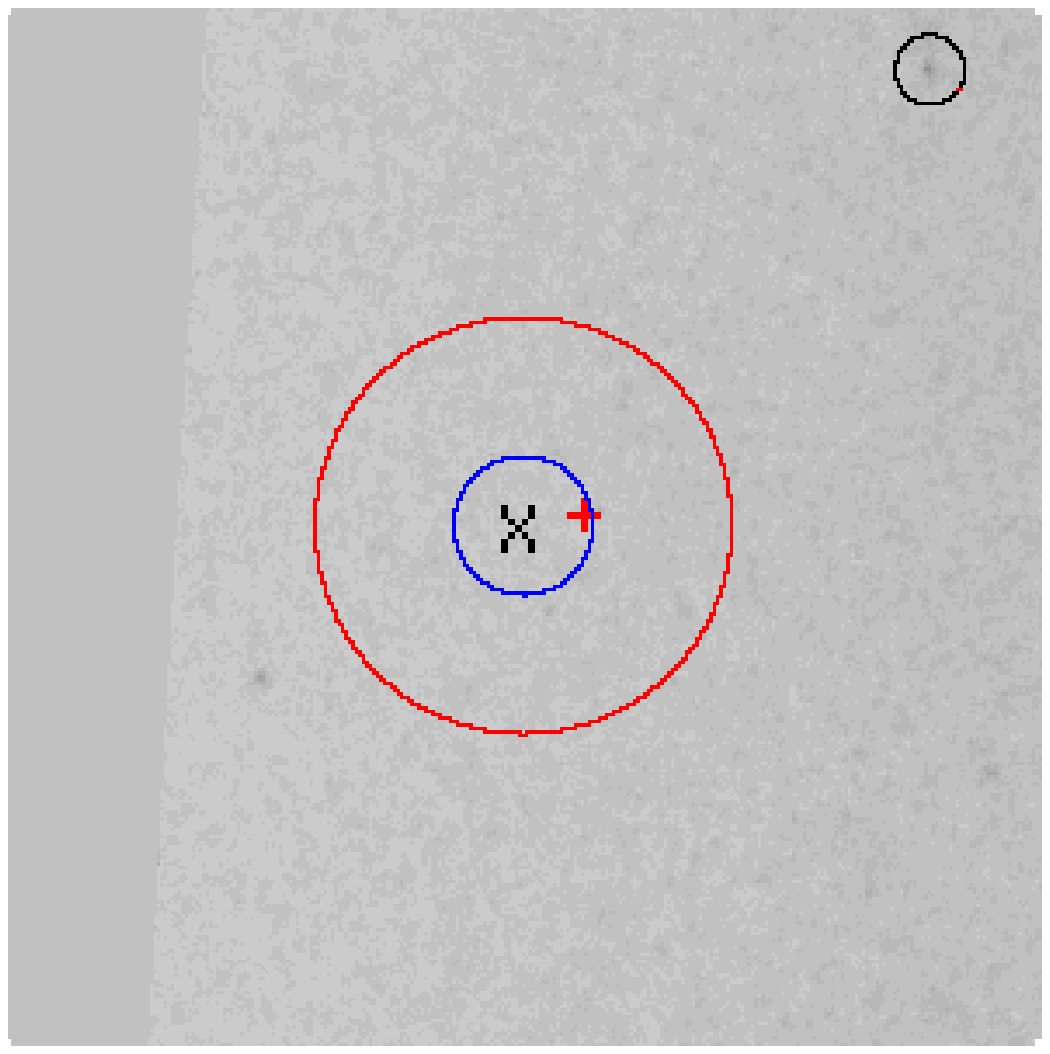}}                
  \subfloat[NGC 4697]{\label{fig:NGC 4697}\includegraphics[width=0.25\textwidth]{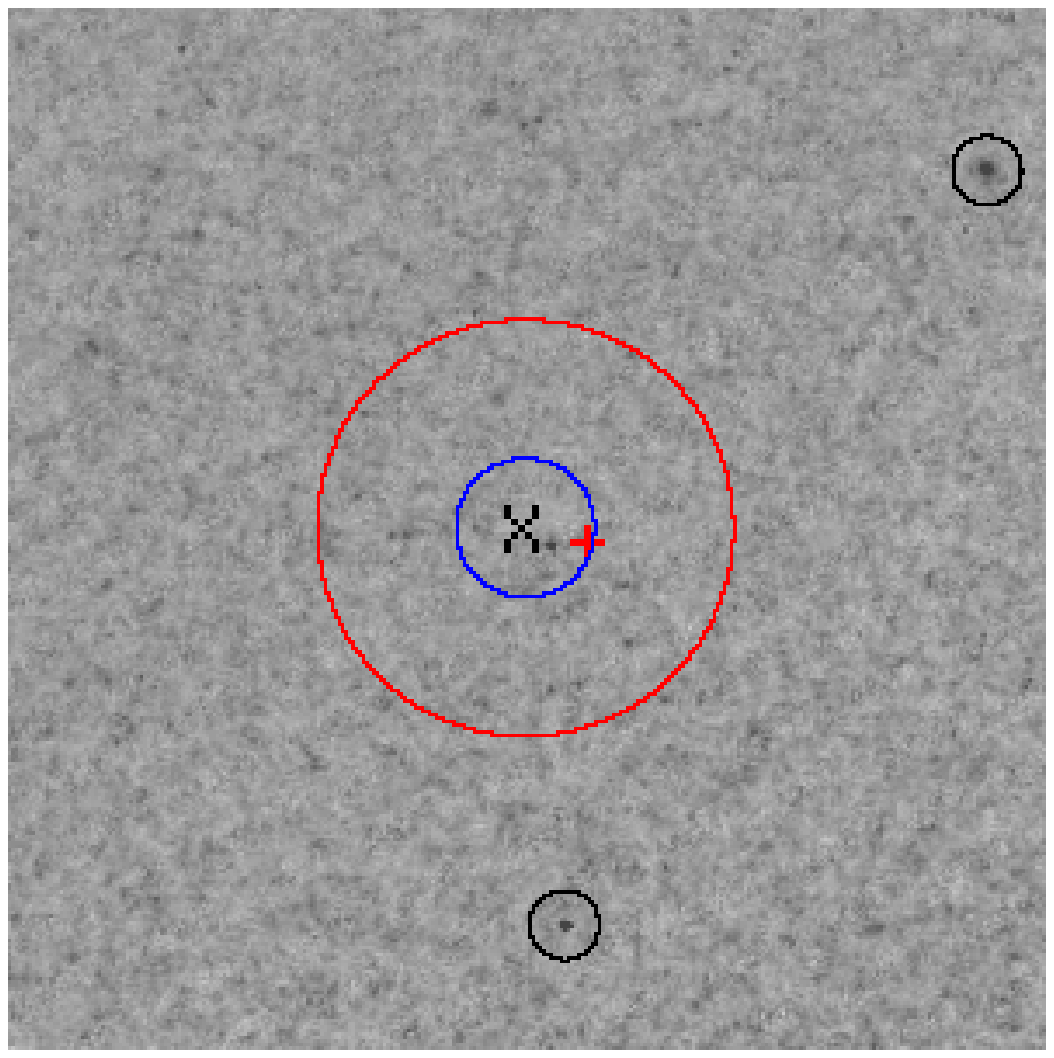}}
  \subfloat[NGC 4649]{\label{fig:NGC 4649}\includegraphics[width=0.25\textwidth]{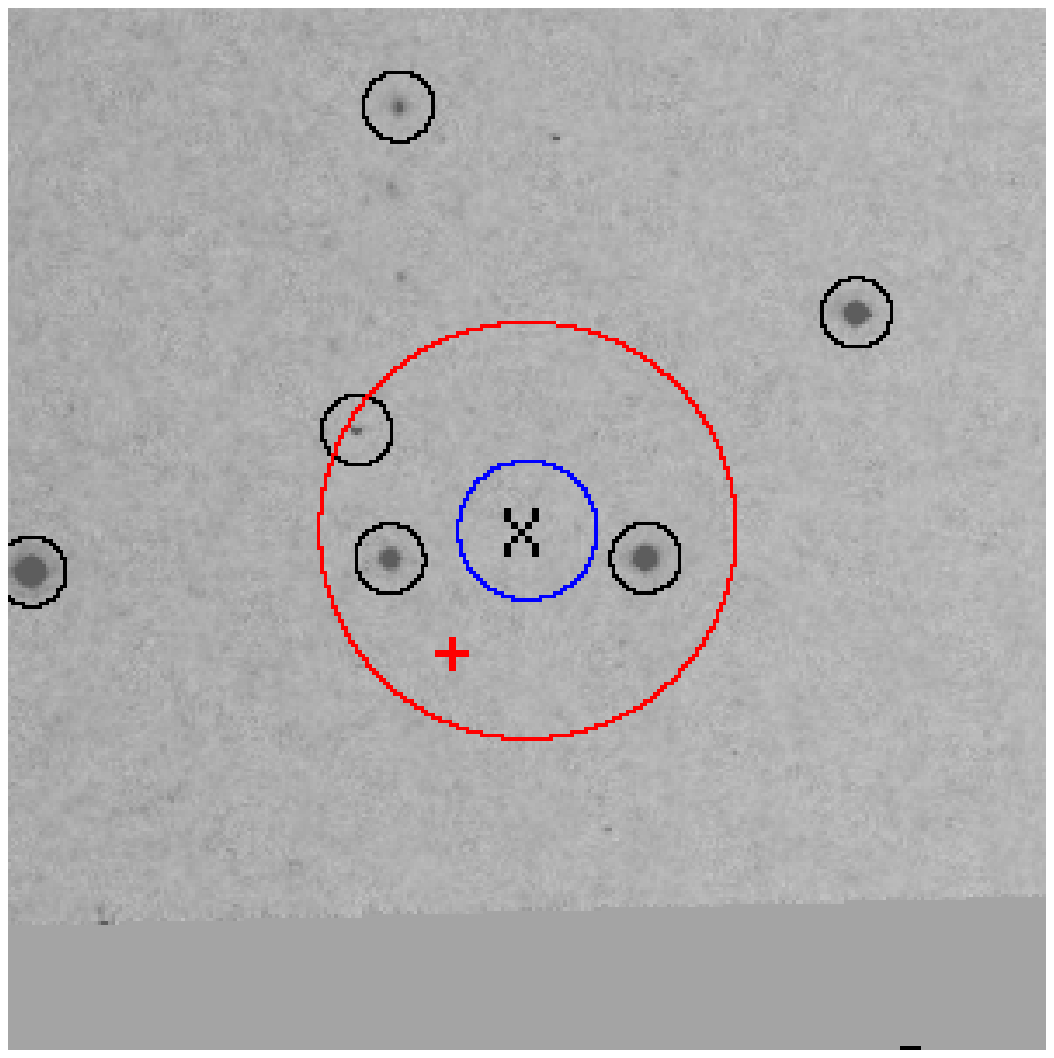}}\\
  \subfloat[NGC 4374]{\label{fig:NGC 4374}\includegraphics[width=0.25\textwidth]{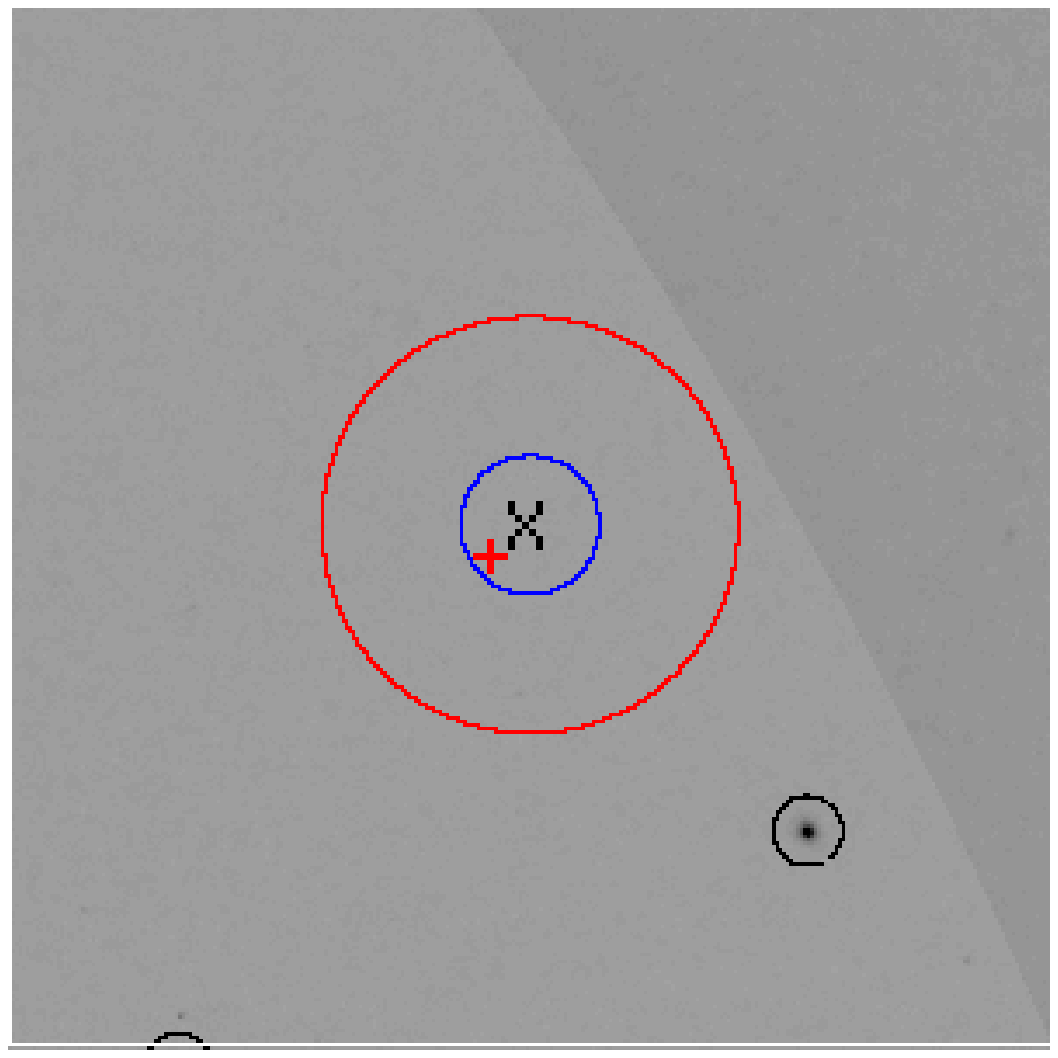}}                
  \subfloat[NGC 1399]{\label{fig:NGC 1399}\includegraphics[width=0.25\textwidth]{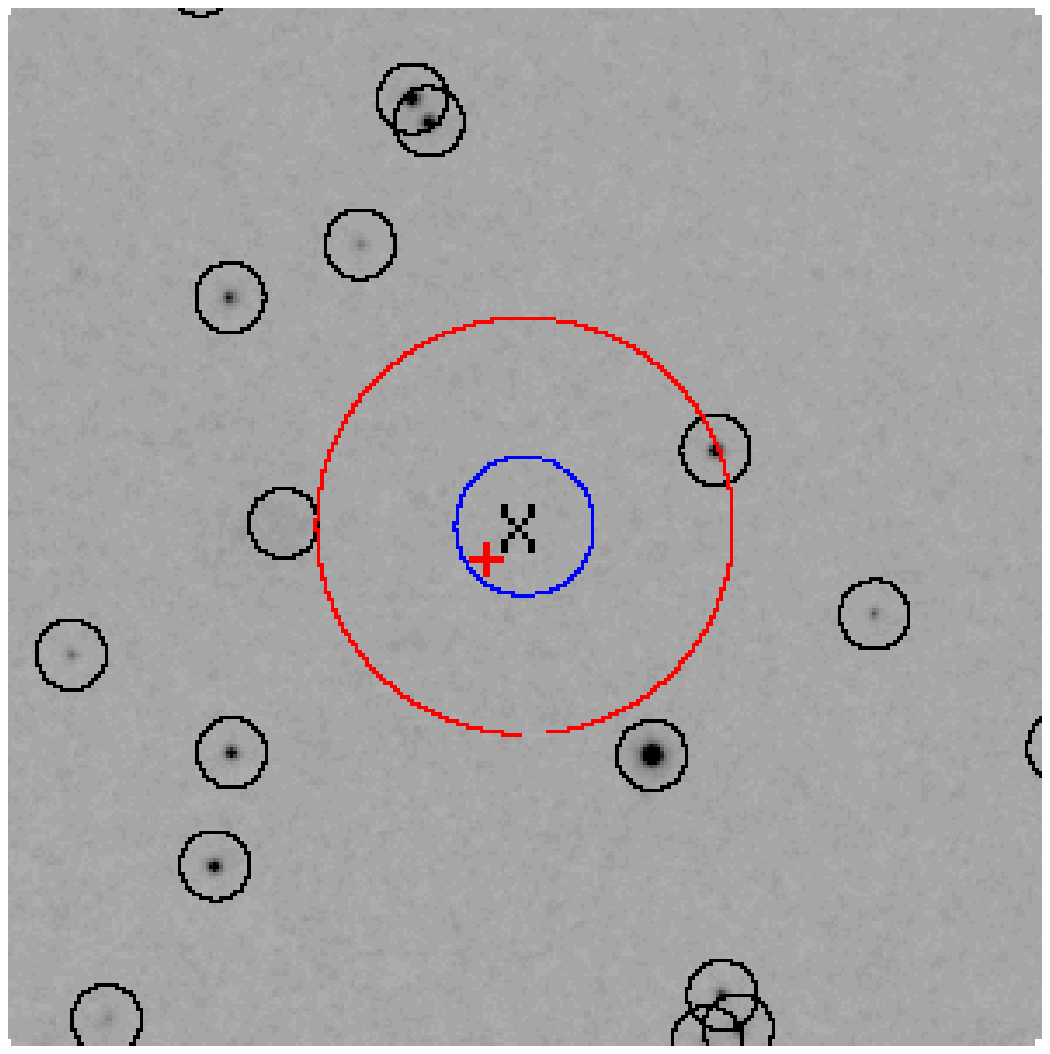}}
  \subfloat[NGC 1316]{\label{fig:NGC 1316}\includegraphics[width=0.25\textwidth]{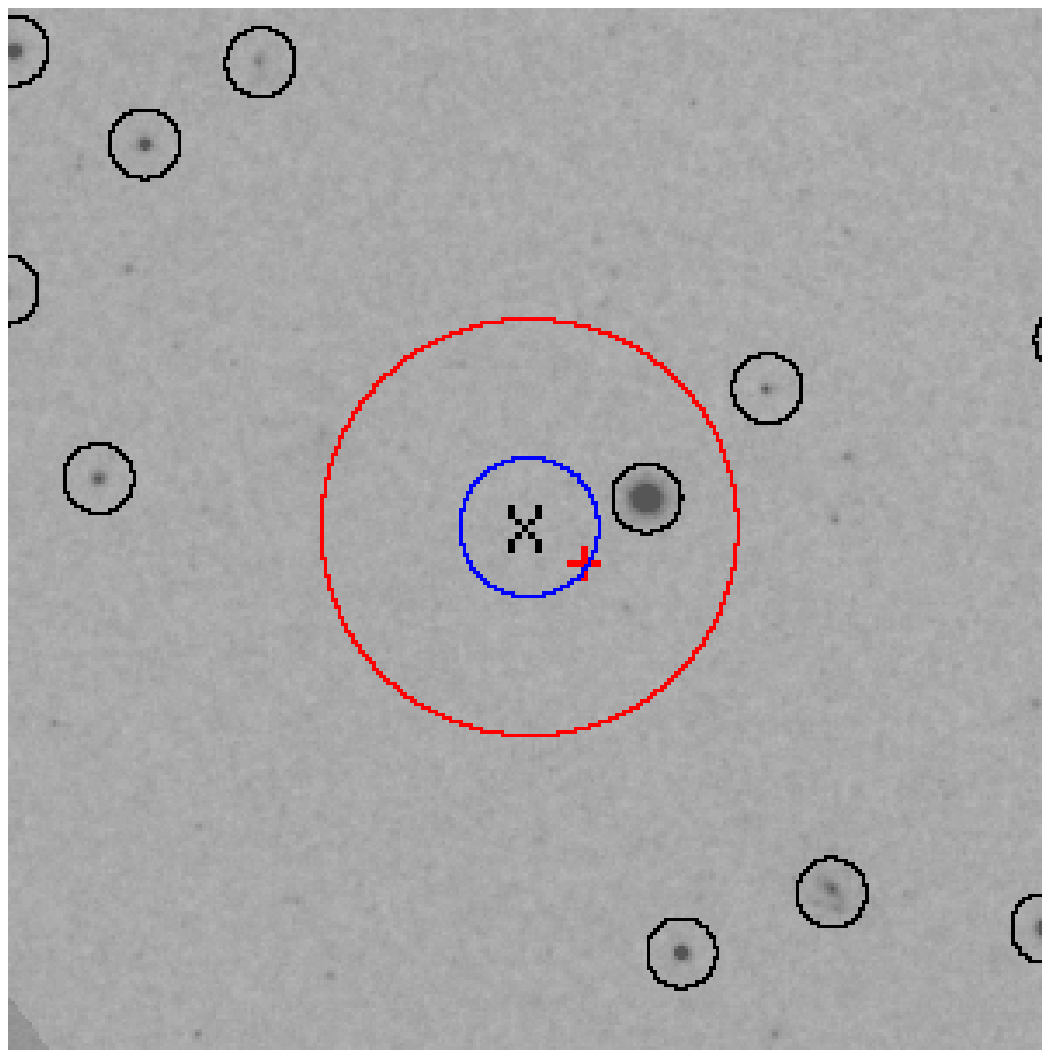}}

\caption{The HST images of the first six X-ray sources (marked X) listed in Table \ref{BHmass}. Overlaid are
$1\arcsec$ and $3\arcsec$ circles centered on the shifted {\it Chandra} positions. The plus signs mark the
original {\it Chandra} positions. Note that for four
of the images, there are no optical sources even within $3\arcsec$ of the X-ray position.
  }
\label{optimg}
\end{center}
\end{figure*}

\section{Observations and Data Reduction}

\cite{Swa04} analyzed  {\it Chandra} observations of 82 nearby galaxies and have identified 
bright X-ray sources in them. Of these we select 13 elliptical galaxies for which 
{ \it HST} observations are available. \cite{Dev07} analyzed a subset of thirty galaxies 
from the \cite{Swa04} sample, and fitted the X-ray points sources with both a power-law 
and a disk black body model, hence obtaining a more conservative and robust estimation 
of their X-ray luminosity. Nine out of the sub-sample analyzed by \cite{Dev07} are 
ellipticals, and for these galaxies we use the X-ray luminosity and coordinates given by 
them for the present analysis. For the remaining four galaxies we use the values quoted 
by \cite{Swa04}. The details of the sample galaxies are given in Table \ref{Sample}.

The optical study was carried out using the images taken with the Advanced Camera for 
Surveys (ACS), and Wide Field and Planetary Camera (WFPC2) that are available in the {\it HST} 
data archive. When observations from both cameras were available, the datasets with the longer 
exposure time, and multiple filters were preferred. For the same combination of camera and 
filters, various images available for different pointings of the same galaxy were analyzed
when available, so as to maximize the number of X-ray sources because for different pointings,
a different set of X-ray sources would fall in the field of view.

\begin{deluxetable} {lllccccr}
\tablewidth{40pc} 
\tablecaption{The Upper limit of Black hole Mass of some of the Optically Dark X-ray Sources} 
\tablehead{
\colhead{Galaxy} & \colhead{${R. A.}(J2000)$} &\colhead{${Decl.}(J2000)$}& \colhead{log $L_{x}$} & \colhead{HST Filter} &\colhead{$F_{\nu}\times10^{-30}$}& \colhead{$F_X/F_O$}& \colhead{$M_U (M_\odot)$}} 
\startdata 
NGC 4486 & 12 30 50.82 & +12 25 02.66 & $39.17^{+0.05}_{-0.04}$ & F475W & 0.409 & 533 & 1244\\ 
NGC 4697 & 12 48 33.20 & -05 47 41.17 & $38.84^{+0.06}_{-0.05}$ & F475W & 0.752 & 243 & 2890 \\
NGC 4649 & 12 43 41.90 & +11 34 33.83 & $38.91^{+0.47}_{-0.11}$ & F475W & 0.402 & 164 & 3073 \\
NGC 4374 & 12 25 01.54 & +12 52 35.59 & $39.10^{+0.67}_{-0.22}$ & F475W & 0.441 & 347 & 3378 \\
NGC 1399 &  3 38 25.92 & -35 27 42.37 & $38.62^{+0.12}_{-0.09}$ & F606W & 0.370 & 228 & 3927 \\
NGC 1316 &  3 22 36.46 & -37 13 24.68 & $38.80^{+0.23}_{-0.08}$ & F475W & 0.449 & 179 & 4780 \\
NGC 1316 &  3 22 35.58 & -37 13 14.10 & $38.78^{+0.40}_{-0.07}$ & F555W & 0.520 & 287 & 6366 \\
NGC 1399 &  3 38 32.33 & -35 26 45.73 & $38.54^{+1.13}_{-0.41}$ & F606W & 0.377 & 186 & 7829 \\
NGC 1399 &  3 38 27.62 & -35 26 48.76 & $39.42^{+0.16}_{-0.14}$ & F606W & 0.766 & 702 & 7829 \\
NGC 4649 & 12 43 34.17 & +11 33 41.93 & $39.04^{+0.10}_{-0.11}$ & F475W & 0.912 & 97 & 8073 \\

\label{BHmass}
\enddata
\tablecomments{(1) Host galaxy name; (2) Right Ascension of shifted position in hours, minutes and seconds; (3) Declination of shifted position in degrees, arcminutes and arcseconds; (4) log 
of X-ray luminosity in $ergs/s$; (5) HST filter for which the upper limit on flux and the black hole mass limit is calculated; (6) Upper limit on Optical 
flux in ergs/s/cm$^2$/Hz; (7) Lower limit on ratio of X-ray to optical flux; (8) $M_U$, Upper limit on black hole mass.}
\end{deluxetable}

Most of the optical sources are too faint to be detected against the dominant galaxy light that 
fills most of the $HST$ images. To enhance the contrast, and aid in the detection of point sources 
in the image, the galaxy light was modeled based on the isophotes obtained using the ellipse task
in { \it IRAF/STSDAS } software. Bright objects, if any, were masked during the fitting. The residual 
image was obtained by subtracting the model image from the observed galaxy image. The object extraction 
was done on the residual image using { \it SEXTRACTOR} with a threshold level of 3 sigma. On visual 
inspection, we find that many of the {\it Chandra} X-ray sources have counterparts in the {\it HST} 
images, within a positional offset of less than a few arc-seconds. This constant offset was applied
for a given galaxy, to match the { \it Chandra} sources to the optical sources in the {\it SEXTRACTOR} 
catalog.  When images are available
in multiple filters, one of the optical images is considered as the reference image, and the necessary
offset is applied to match it to the X-ray co-ordinates. The images in other optical filters are then
aligned to the reference image using { \it geomap} and { \it geotran} tasks in IRAF. While a more
detailed report on the nature of sources with optical counterparts will be presented later, in
this work, we concentrate on those sources for which no optical counterpart was detected.

X-ray sources which did not have an optical counterpart (at the 3$-$sigma level) 
within $1\arcsec$ of their shifted positions, are
termed as optically ``dark'' sources.
Six examples of such sources
are shown in Figure \ref{optimg}, where the $1\arcsec$ and $3\arcsec$ circles centered on the X-ray co-ordinates are 
overlaid on the observed optical image. Note that for four of these images, there are no optical source within
$3\arcsec$ of the X-ray position.  Having 
identified such optically "dark" X-ray sources, we estimate the upper limit on 
their optical flux based on the 3$-$sigma threshold at that position.

\section{Optically dark X-ray sources}

These optically dark sources are X-ray bright compared to their
optical emission and hence are not foreground stars. This can be further quantified by 
estimating the X-ray-to-optical flux ratio $
\log\left(f_X/f_O\right)$ where $f_X$ is the unabsorbed flux in the $0.3-8{\rm~keV}$ band and $f_O$ is the flux in an optical band. This ratio ranges from $0.1$ to $50$ for AGNs
including BL~Lacs when $f_X$ is in the $0.3-3.5{\rm~keV}$ band and the V-band magnitude is used 
\citep{Sto91}. In contrast, the estimated lower limit for the optically dark sources
in the sample is significantly larger. This is illustrated in Table (\ref{BHmass}) 
where the ratio is given for ten sources. We, therefore, conclude that these sources are  not
background AGNs.

Thus, these sources are most likely to be bright X-ray binaries (or at
least accreting systems) within the galaxy. An accretion disk around a
compact object should also produce optical emission whose flux can be
estimated as follows. In the standard accretion disk theory
\citep{Sha73}, the effective temperature profile as a function of
radius $R$, of an accretion disk around a black hole with mass $M$ and
accretion rate $\dot M$, is given by $
\sigma T^4 (R) = \frac{3}{8 \pi} \frac{GM\dot M}{R^3} \delta (R) $
where $\delta (R) = 1 - (6GM/Rc^2)^{1/2}$. 
The observed flux from the disk at a frequency $\nu$ is then given by the
integrated sum of the black body emission over all radii,
\begin{equation}
F_\nu = \frac{{\hbox{cos}}{i}}{D^2}\int^{R_{out}}_{R_{in}} B_\nu (\nu,T(R))\; 2 \pi R dR
\label{Eqn.Fv}
\end{equation}
where $B_\nu (\nu,T(R))$ is the blackbody intensity, $R_{out}$ and $R_{in}= 6GM/c^2$ are
outer and inner radii of the disk, $i$ is the inclination angle of the disk and
$D$ is the distance to the source. Assuming that most of the contribution to $F_\nu$ arises
from regions in the disk that are far away from the inner and outer radii, the expected
observed flux can be written as
\begin{eqnarray}
F_\nu & \sim & 7 \times 10^{-31} \;\; \hbox{ergs s}^{-1}\hbox{cm}^{-2}\hbox{Hz}^{-1}  (\frac{\lambda}{5000 A})^{-1/3} (\frac{\eta}{0.1})^{-2/3} \nonumber \\
 & & \times (\frac{L_x}{10^{39} \hbox{ergs}\;{s}^{-1}})^{2/3} (\frac{D}{10 \hbox{Mpc}})^{-2} (\frac{M}{1000 M_\odot})^{2/3} 
\end{eqnarray}
where $\lambda$ is the wavelength and $\eta = L_x/\dot M c^2$ is the radiative efficiency of the 
accreting system and cos$i$ is taken to be $0.5$.

For  optically dark ULX, the predicted accretion flux should be less than the measured upper limit 
$F_{\nu, max}$. Thus one can estimate an upper limit on the black hole mass as
\begin{eqnarray}
M_U & <  &  1000\;\; M_\odot \; (\frac{F_{\nu, max}}{7 \times 10^{-31} \;\; \hbox{ergs s}^{-1}\hbox{cm}^{-2}\hbox{Hz}^{-1}})^{3/2} \nonumber \\
 & & \times (\frac{\lambda}{5000 A})^{1/2} (\frac{\eta}{0.1}) (\frac{L_x}{10^{39} \hbox{ergs}\;{s}^{-1}})^{-1}  (\frac{D}{10 \hbox{Mpc}})^{3} 
\label{M_U}
\end{eqnarray}
For each dark X-ray source in our sample, and for all available filters, we estimate this upper limit 
on the black hole mass. We use the integration (Eqn \ref{Eqn.Fv}) to evaluate the upper limit, rather 
than the approximation Eqn. (\ref{M_U}) i.e. we take into account the effect of the inner boundary 
condition ($\delta (R) = 1 - (6GM/Rc^2)^{1/2}$) on the temperature profile. The difference in the upper 
limit obtained is marginal ($< 20$\%). We assume a standard radiative efficiency of $\eta = 0.1$ and 
cos$i = 0.5$. To obtain a more conservative upper limit, the one-sigma lower value of the X-ray 
luminosity are used $L_x - \Delta L_x$. In Table \ref{BHmass}, we list the ten best cases in ascending 
order of black hole mass limit $M_U$. For the other sources, $M_U > 10000 M_\odot$ and hence is not 
a significant constraint. The best case is for the ULX in NGC 4486, for which $M_U = 1244 M_\odot$. 

X-ray irradiation of the outer disk may increase the local temperature there and the disk may emit a 
larger optical emission. We have estimated this effect using the formalism given in the appendix of 
\cite{Vrt90} and find that X-ray irradiation is not important for the constraint obtained here.

\section{Discussion}

Optically dark X-ray sources cannot be foreground stars or background 
AGN, otherwise their optical emission would be significantly higher than what is detected. Hence, these
are a clean sample of sources within the host galaxies, which are probably accreting black hole systems.
The optical emission from a standard accretion disk scales as mass of the black hole $M^{2/3}$ and 
hence the non-detection of optical emission imposes an upper limit on the black holes mass $M_U$. For 
ten of the sources $M_U < 10000 M_\odot$. For a source in NGC 4486 with an X-ray 
luminosity clearly exceeding $10^{39}$ ergs/s (and therefore a bona-fide ULX by definition), the 
estimated black hole mass is smaller than $1244 M_\odot$. This is two orders of magnitude smaller 
than the constraint obtained from dynamical friction, which is $ 10^5 M_\odot$. 

These sources with black hole mass, $M_U < 5000 M_\odot$ cannot be accreting systems with massive 
black holes residing in star clusters, or in the nuclei of merged satellite galaxies. For typical
low-luminosity dwarf galaxies \citep[$M_{B} \sim -8.0$;][]{Mat98}, such an optical counterpart would be
easily detected, given that our 3$-$sigma limits on the $HST$ images are much fainter. Even a compact 
nucleus of a merged dwarf galaxy \citep[$M_{B} \sim -7.0$;][]{Lot04} would have been easily 
identified. If they are binary systems, their companion cannot be a massive O star as such a star 
would have been detected in the optical image. Assuming an O star, with $M_{B} \sim -5.5$,
we find that the possibility of such a companion can be ruled out in all cases for which
$M_U < 5000 M_\odot$ (Table \ref{BHmass}).

In all of the above arguments, we have ignored the effect of dust
obscuration in the host galaxies, because we are only considering elliptical galaxies in the 
present work. Although some ellipticals are known to have dust lanes, and nuclear rings in their
centers, the X-ray sources we are considering here are distributed at fairly large radial distance
from the center to be significantly affected by dust.

Even for the best cases of optically dark X-ray sources 
presented in Table (\ref{BHmass}), the range of black hole mass allowed is still large as the source could be 
a $\simless 1000 M\odot$ intermediate mass black hole, or a few solar mass object emitting at 
super-Eddington luminosities.
The brightest X-ray sources in the sample have a luminosity of a few times $10^{39}$ ergs/s. This 
is unfortunate, since a dark source with  luminosity $ > 10^{40}$ would have provided an order of 
magnitude better constraint on the black hole mass. Since the black hole mass upper limit $M_U 
\propto D^3$ a bright X-ray source in a more nearby galaxy ($D \sim 2 Mpc$) would have also provided 
significantly better constraints. A systematic search for such sources in very nearby galaxies may 
indeed prove fruitful. Another point to note is that bright X-ray sources in these galaxies are 
known to be variable in X-rays. Our analysis in this work, implicitly assumes that the X-ray 
luminosity observed through a single {\it Chandra} observation, represents an average luminosity 
which is used to derive an average accretion rate $\dot M = L_X/(\eta c^2)$ which in turn is used 
to estimate the upper limit on  the black hole mass (Eqn \ref{M_U}). This assumption is required 
because the expected optical emission arises from the outer part of the disk and the local accretion 
rate there may be different than the one in the inner region which produces the X-rays. Any accretion  
rate fluctuation in the outer disk will be transfered along the disk on the viscous time-scale which 
could be significantly longer than a day. Thus, in principle one needs to ascertain the average X-ray 
luminosity of a source, and using a single very bright, but rare, X-ray observation of the source 
will not represent the average accretion rate.

A systematic and comprehensive multi-wavelength study (using also other bands like infra-red and 
radio), along with X-ray variability studies, can shed further light on the nature of these sources.

\acknowledgements

VJ, KJ, CDR and BRSB would like to thank the IUCAA visitors program
and and UGC Special assistance program. This
work has been partially funded from the ISRO-RESPOND programme. The authors would like to
thank Phil Charles for useful discussions.

\end{document}